\documentclass[a4paper,11pt]{article}

\usepackage{jheppub}

\usepackage{amssymb,amsthm,mathtools}
\usepackage{microtype}
\usepackage{tikz}
\usepackage{tikz,graphicx}
\usepackage{booktabs}
\usetikzlibrary{arrows.meta}
\hypersetup{
  hidelinks,
  pdftitle={Bounds on the Diagonal Degree in the KLT Construction},
  pdfsubject={}
}

\newtheorem{theorem}{Theorem}[section]

\theoremstyle{definition}
\newtheorem{example}[theorem]{Example}

\newcommand{\rank}{\operatorname{rank}}
\newcommand{\supp}{\operatorname{supp}}

\newcommand{\Sn}{\mathfrak{S}}
\newcommand{\cA}{\mathcal{A}}
\newcommand{\cB}{\mathcal{B}}

\newcommand{\one}{\mathbf{1}}
\newcommand{\DeltaDiag}{\Delta}

\title{\boldmath Bounds on the Diagonal Degree of the Biadjoint Scalar Amplitude Matrix}

\author[a]{Freddy Cachazo}
\author[b]{and Rajbir Longia}

\affiliation[a]{Perimeter Institute for Theoretical Physics,
Waterloo, ON N2L 2Y5, Canada}

\affiliation[b]{Department of Combinatorics \& Optimization,
University of Waterloo, Waterloo, ON N2L 3G1, Canada}

\emailAdd{fcachazo@pitp.ca}
\emailAdd{rlongia@uwaterloo.ca}

\abstract{ Biadjoint scalar theory partial amplitudes, $m_n(\alpha,\beta)$, define an $(n-1)!/2\times (n-1)!/2$ matrix, $M_n$, with rows and columns labeled by orderings. The CHY construction of $M_n$ implies that its rank is $(n-3)!$. Invertible submatrices of maximal rank enter in the KLT construction relating graviton amplitudes to gluon amplitudes. Motivated by finding the KLT formula with the smallest number of terms, one of the authors posed the question of whether there is a critical $n_c$ such that for all $n>n_c$ there exists a diagonal submatrix of $M_n$ of rank $(n-3)!$. In this work we obtain bounds on the size of the largest non-singular diagonal submatrix of $M_n$, which we call its diagonal degree $\Delta_n$, and we answer the question in the negative. We prove that the diagonal degree is exponentially suppressed relative to the CHY rank: $\Delta_n/(n-3)! = \exp(-\Theta (n))$ as $n\to \infty$. More explicitly, we establish the bounds $\exp(-1.5n)\leq\Delta_n/(n-3)!\leq \exp \left[ -0.01n+O(1)\right]$.
}

\begin{document}

\maketitle

\section{Introduction}

The Kawai-Lewellen-Tye (KLT) relation \cite{KLT} discovered in the context of string theory in the 80s connects scattering amplitudes of $n$ gravitons to linear combinations of products of two partial amplitudes of $n$ gluons. The precise linear combination is controlled by the inverse of a matrix of biadjoint scalar partial amplitudes. 

The $U(N)\times U(\tilde N)$ biadjoint scalar theory admits a double color decomposition and its partial amplitudes depend on two orderings. Orderings are permutations of $(1,2,\ldots, n)$ modulo the action of the dihedral group $D_n$. Partial amplitudes are traditionally denoted as $m_n(\alpha,\beta)$, with $\alpha$ and $\beta$ two orderings. These amplitudes can be arranged in an $(n-1)!/2 \times (n-1)!/2$ matrix with rows and columns labeled by orderings. Let us denote such a matrix as $M_n$.

The matrix $M_n$ is known to have rank $(n-3)!$ (see section 3 for a review). The KLT construction uses the inverse of any non-singular $(n-3)!\times (n-3)!$ submatrix of $M_n$. Let us denote by $m^{\rm KLT}_n$ a generic rank $(n-3)!$ submatrix of $M_n$. 

Not all matrices $m^{\rm KLT}_n$ have the same support; some are more sparse than others. Clearly, the complexity of the KLT construction depends on how sparse $(m^{\rm KLT}_n)^{-1}$ is. Generically, $(m^{\rm KLT}_n)^{-1}$ is a very dense matrix and the KLT formula requires close to $(n-3)!^2$ products of gluon amplitudes \cite{FCQ}.  

The search for the KLT formula with the smallest number of terms motivated one of the authors to try and find diagonal matrices $m^{\rm KLT}_n$ \cite{FCQ}. For $n<5$ the question is trivial as any $(n-3)!\times (n-3)!$ matrix is diagonal. For $n=5$, such a $2\times 2$ diagonal matrix exists, but for $n=6$ the largest non-singular diagonal submatrix of $M_n$ is only $4\times 4$. For $n=7$ the largest known non-singular diagonal submatrix has size $14\times 14$. While this seems to indicate that finding a diagonal matrix with rank $(n-3)!$ is impossible for $n>5$, the density of the matrices $M_n$ can be shown to decrease exponentially with $n$ for large $n$; in particular the exponential decay starts to dominate the polynomial factor for $n>16$. This prompted the question of whether there exists a critical value $n_c$ such that for $n>n_c$, there are  diagonal submatrices $m_n$ with maximal rank, i.e. rank $(n-3)!$. 

In this work we answer this question in the negative and provide asymptotic upper and lower bounds for the ranks of such diagonal submatrices of $M_n$. 

Let us denote by $\Delta_n$ the rank of the largest non-singular diagonal submatrix of $M_n$. We prove that 
\begin{equation}
    \frac{\Delta_n}{(n-3)!} = \exp\left[ - \Theta(n) \right].
\end{equation}
More explicitly, 
\begin{equation}
    \exp(-1.5n)\leq\Delta_n/(n-3)!\leq \exp \left[ -0.01n+O(1)\right].
\end{equation}
The upper bound $O\left( e^{-0.01 n}\right)$ is obtained by a construction using balanced tripartitions to build a matrix with the same support as $M_n$ but with entries made out of products of smaller biadjoint amplitudes. The balanced condition keeps the ranks of the smaller amplitudes small and this leads to the upper bound after applying a recursive argument.

The lower bound $\Omega\left( e^{-1.5 n}\right)$ is obtained by starting with two  explicit sets of $(n-3)!$ orderings that produce a block diagonal matrix $m$ with blocks of size $\lfloor (n-3)/2\rfloor !\lceil (n-3)/2\rceil !$. The bound is then obtained by recursion of the construction on each of the blocks.

This work is organized as follows: In section 2, we define the problem in terms of scattering amplitudes and then in a purely combinatorial form. As it turns out, the former is used for the upper bound while the latter for the lower bound. In section 3, we review the proof that the rank of the matrix $M_n$ is $(n-3)!$. In section 4, we present the derivation of the upper bound. In section 5, we explain how the lower bound was obtained. In section 6, we end with discussions and future directions.

\section{Defining the Diagonal Degree and Its Physics Motivation}

Tree-level scattering amplitudes of $n$ gluons in $SU(N)$ Yang-Mills theory can be computed using Feynman diagrams that involve cubic and quartic vertices. Moreover the interaction vertices depend on the adjoint index ${\tt a}\in \{1,2,\ldots ,N^2-1\}$ of the interacting particles. Each external particle is defined by a momentum vector, a polarization vector, and its color index, i.e., $\{ k_a^\mu,\epsilon_a^\mu, {\tt a}_a \}$. Yang-Mills amplitudes admit a color decomposition into partial amplitudes (see \cite{ManganoParke} for a review),
\begin{equation}
    {\cal A}^{\rm gluons}_n(\{ k_a,\epsilon_a, {\tt a}_a \}) =\sum_{\alpha\in S_n/\mathbb{Z}_n}{\rm Tr}\left(T^{{\tt a}_{\alpha(1)}}T^{{\tt a}_{\alpha(2)}}\cdots T^{{\tt a}_{\alpha(n)}}\right) A_n(\alpha_1,\alpha_2,\ldots ,\alpha_n).
\end{equation}
Here partial amplitudes $A_n(\alpha_1,\alpha_2,\ldots ,\alpha_n)$ (or $A_n(\alpha)$ for short) are computed by summing over Feynman diagrams which are planar with respect to the $\alpha$ ordering and contain no $SU(N)$ information. All color information is in the traces of products of generators, $T^{\tt a}$, in the fundamental representation. 

Tree-level scattering amplitudes of $n$ gravitons are computed by writing the metric $g_{\mu\nu}=\eta_{\mu\nu}+\kappa h_{\mu\nu}$ around flat space and expanding the Einstein-Hilbert action to $n^{\rm th}$ order to read out the interaction vertices. The corresponding Feynman diagrams are trees with vertices allowed to have all degrees up to $n$ and each external particle is defined by a momentum vector and a polarization tensor, $\{ k_a^\mu ,\epsilon_a^{\mu\nu}\}$. Let us denote amplitudes of gravitons by ${\cal A}^{\rm gravitons}_n(\{ k_a^\mu,\epsilon_a^{\mu\nu} \})$. 

The last physics ingredient consists of the tree-level scattering amplitudes of an $SU(N)\times SU(\tilde N)$ biadjoint massless cubic scalar theory. These amplitudes are computed using only cubic trees but vertices involve structure constants of two groups. Each external particle is defined by $\{k_a^\mu , {\tt a}_a, \tilde{\tt a}_a\}$. These amplitudes admit a double color decomposition
\begin{equation}
    {\cal A}^{\rm scalars}_n(\{ k_a,{\tt a}_a,\tilde{\tt a}_a \}) =\sum_{\alpha, \tilde{\alpha}\in S_n/\mathbb{Z}_n}{\rm Tr}\left(\alpha \right){\rm Tr}\left(\tilde\alpha \right) m_n(\alpha,\tilde\alpha)
\end{equation}
where ${\rm Tr}\left(\alpha \right)$ is a shorthand notation for the trace of the generators in the $\alpha$ order and $m_n(\alpha,\tilde\alpha)$ is the double partial amplitude of scalars computed as the sum over Feynman diagrams corresponding to cubic trees and planar with respect to $\alpha$ and $\tilde\alpha$.  

The Kawai-Lewellen-Tye (KLT) relation, originally discovered in string theory and translated into quantum field theory language in \cite{BernKLT}, is a surprising explicit relation among graviton and gluon amplitudes,
\begin{equation}
\label{eq:klt1}
   {\cal A}^{\rm gravitons}_n(\{ k_a^\mu,\epsilon_a^{\mu\nu}\}) = \sum_{\alpha\in A, \beta\in B}A_n(\alpha)S[\alpha |\beta ]A_n(\beta). 
\end{equation}
Here $A$ and $B$ are sets of $(n-3)!$ orderings. $A$ and $B$ do not have to be the same and the choices coming from string theory define a $(n-3)!\times (n-3)!$ matrix $S[\alpha | \beta]$, known as the {\it momentum kernel}, which is invertible. $S[\alpha | \beta]$ is a rational function of kinematic invariants involving only the momenta $k^\mu_a$ in the Lorentz invariant combinations $s_{ab}:=2k_a\cdot k_b$. We review its explicit form for some choices of $A$ and $B$ in section 5.


Here we review another surprising connection that was discovered using the Cachazo-He-Yuan (CHY) formalism (see section 3) relating the momentum kernel to biadjoint scalar amplitudes. Define $M_n$ to be the $(n-1)!/2\times (n-1)!/2$ matrix of double partial biadjoint scalar amplitudes, so
\begin{equation}
    (M_n)_{\alpha, \beta} := m_n(\alpha, \beta).
\end{equation}

In the next section we review the CHY formulation and the proof that ${\rm rank}(M_n)=(n-3)!$. Let $m_n^{\rm KLT}$ denote any non-singular $(n-3)!\times (n-3)!$ submatrix of $M_n$. Let $A$ and $B$ be the sets of orderings that label the rows and columns of $M_n$ used to construct $m_n^{\rm KLT}$. The following is the relationship between the momentum kernel and the matrix of biadjoint amplitudes \cite{Cachazo:2013gna}
%
%
\begin{equation}
    S[\alpha | \beta] = (m_n^{\rm KLT})^{-1}_{\beta,\alpha}, \quad \alpha\in A,~\beta\in B.
\end{equation}
Generic choices of $A$ and $B$ lead to very dense matrices $S[\alpha | \beta]$ and hence ${\cal O}\left( (n-3)!^2\right)$ terms are needed in the KLT relation \eqref{eq:klt1}. A natural question is how sparse $S[\alpha | \beta]$ can be chosen to be. 

In this work, we do not directly study $S[\alpha | \beta]$; instead we study $m_n(\alpha , \beta)$. The correlation of how sparse a matrix and its inverse are is only straightforward when they are block diagonal. This motivates the following definition.

The {\it diagonal degree}, $\Delta_n$, is the size of the largest non-singular diagonal submatrix of $M_n$. 

The first non-trivial example is for $n=5$. Here $\Delta_5=2$ and can be explicitly realized by choosing
\begin{align}
    A = B =  \{ (1,2,3,4,5),(1,3,5,2,4) \}.
\end{align}
The first case in which the diagonal degree is less than $(n-3)!$ is for $n=6$. An exhaustive computer search resulted in $\Delta_6=4$ \cite{FCQ}, with an explicit realization given by 
\begin{align}
    A & =  \{ (1,2,3,4,5,6),(1,2,6,3,4,5),(1,5,3,4,2,6),(1,5,4,2,3,6) \},  \\
    B & =  \{ (1,3,2,5,6,4),(1,2,5,3,6,4),(1,3,5,2,6,4),(1,3,6,5,2,4) \}.
\end{align}

In the next section we review how the upper bound $\Delta_n \leq (n-3)!$ is derived using the CHY formalism.

\section{Review of the CHY Bound}

The CHY formulation expresses double partial amplitudes of the massless biadjoint scalar theory as follows \cite{Cachazo:2013hca,Cachazo:2013iea},
\begin{equation}
    m_n(\alpha, \beta) = \int \frac{d^n\sigma}{{\rm vol}\, SL(2,{\mathbb C})}\sideset{}{'}\prod_a \delta\left(\sum_{b\neq a} \frac{s_{ab}}{\sigma_a-\sigma_b}\right) {\rm PT}(\alpha){\rm PT}(\beta).
\end{equation}
The integrand is invariant under the standard action of $SL(2,{\mathbb C})$ on $\sigma_a$. The factor ${\rm vol}\, SL(2,{\mathbb C})$ is there to indicate that a gauge fixing is needed. This procedure fixes three variables and removes three delta functions. This is indicated by the prime in the product. Also, the so-called Parke-Taylor factors are defined by
\begin{equation}
    {\rm PT}(\alpha) :=\frac{1}{(\sigma_{\alpha_1}-\sigma_{\alpha_2})(\sigma_{\alpha_2}-\sigma_{\alpha_3})\cdots (\sigma_{\alpha_n}-\sigma_{\alpha_1})}.
\end{equation}
Let us fix the $SL(2,{\mathbb C})$ invariance by fixing integration variables $\{ n-2,n-1,n\}$ to $\sigma_{n-2}^*,\sigma_{n-1}^*,\sigma_n^*$ and removing equations $\{n-2,n-1,n\}$. This gives rise to 
\begin{equation}
    m_n(\alpha, \beta) = \left. \int d^{n-3}\sigma |n-2,n-1,n|^2\prod_{a=1}^{n-3} \delta\left(\sum_{b\neq a} \frac{s_{ab}}{\sigma_a-\sigma_b}\right) {\rm PT}(\alpha){\rm PT}(\beta)\right|_{\sigma_{c}
    \rightarrow\sigma_{c}^*,\, n-2\leq c\leq n},
\end{equation}
with $|a,b,c|:=(\sigma_a-\sigma_b)(\sigma_a-\sigma_c)(\sigma_b-\sigma_c)$. One can prove that the value of the integral is independent of the choices made.

The equations imposed by the delta functions are known as the {\it scattering equations} and have $(n-3)!$ solutions for generic values of $s_{ab}$ satisfying $\sum_b s_{ab}=0$ for all $a\in [n]$ \cite{Cachazo:2013gna}. This can easily be proven by noting that the equations are the vanishing gradient conditions for a Morse function on ${\cal M}_{0,n}$ and hence the number of solutions computes its signed Euler characteristic, ultimately giving $(n-3)!$ \cite{Mizera:2019gea}.
Explicitly evaluating the CHY integral gives rise to
\begin{equation}
    m_n(\alpha, \beta) =\sum_{I=1}^{(n-3)!} \left.\frac{{\rm PT}(\alpha){\rm PT}(\beta)}{\det'\Phi }\right|_{\sigma_a=\sigma_a^{(I)}}, \quad \Phi_{ab}:=\partial_{\sigma_b}E_a, \quad E_a:=\sum_{b\neq a} \frac{s_{ab}}{\sigma_a-\sigma_b}.
\end{equation}
Here $\sigma_a^{(I)}$ stands for the $I^{\rm th}$ solution and ${\det'\Phi }$ for the determinant of the submatrix of $\Phi$ obtained by removing the $\{ n-2,n-1,n\}$ rows and columns and dividing by $|n-2,n-1,n|^2$. 

Let us introduce an $(n-3)!\times (n-1)!/2$ matrix $V$ and an $(n-3)!\times (n-3)!$ matrix $Q$ as follows 
\begin{equation}
    V_{I,\alpha}:=\left. {\rm PT}(\alpha)\right|_{\sigma_a=\sigma_a^{(I)}}, \quad Q_{IJ}:=\frac{\delta_{IJ}}{(\det'\Phi)_{\sigma_a=\sigma_a^{(I)}}}.
\end{equation}
Now one can write 
\begin{equation}
    m_n(\alpha, \beta) =\sum_{I,J=1}^{(n-3)!}V_{I,\alpha}Q_{IJ}V_{J,\beta}, \quad {\rm hence}\quad M_n = V^T\, Q\, V.
\end{equation}
This immediately gives 
\begin{equation}
    \Delta_n \leq \rank M_n \leq (n-3)!.
\end{equation}
For general kinematics, the scattering-equation solutions are nondegenerate, so $Q$ is invertible, while KLT orthogonality \cite{Cachazo:2013gna} implies that $V$ has full row rank. Consequently, $\rank M_n = (n-3)!$.


The fact that $m_n(\alpha, \beta)$, as defined in this section, computes the double partial amplitude of the biadjoint scalar theory was conjectured in \cite{Cachazo:2013iea} and proven by Dolan and Goddard in \cite{Dolan:2013isa}.

Let us review how the biadjoint amplitude is computed using Feynman diagrams. Such a description is useful in combinatorial arguments in the rest of this work.

The first ingredient is the kinematic invariant associated with a subset  $I\subset [n]$ with $2\leq |I| \leq n-2$, 
$$s_I:= \sum_{\{a,b\} \in \binom{I}{2}}s_{ab}.$$
Note that $s_I = s_{[n]\setminus I}$ is a consequence of momentum conservation\footnote{This is the condition that $\sum_{a=1}^n k^\mu_a=0$, which implies that $\sum_{b=1}^ns_{ab}=0$ for all $a\in [n]$.}. The value of the Feynman diagram associated with a tree $T$ is then
\begin{equation}
    \Gamma (T) = \prod_{e\in E^{\rm int}(T)} \frac{1}{s_{I_e}},
\end{equation}
where $E^{\rm int}(T)$ is the set of internal edges of $T$ and $I_e\cup ([n]\setminus I_e)$ is the partition of $[n]$ associated to $e$. 

Let $\Omega (\alpha)$ be the set of all cubic  trees that are planar with respect to $\alpha$. The biadjoint scalar amplitude is then given by
\begin{equation}
\label{eq:biadj-tree}
    m_n(\alpha,\beta) = (-1)^{w(\alpha,\beta)}\sum_{T\in \Omega(\alpha)\cap \Omega(\beta)} \Gamma(T),
\end{equation}
where $w(\alpha, \beta)$ is a winding number which is important in order to show that the matrix $M_n$ has rank $(n-3)!$ when using the expression in terms of trees. In this work, we use trees for combinatorial arguments only, while the ranks are computed using the CHY formulation.
The tree expansion gives the following criterion for nonvanishing:
\begin{equation}
\label{eq:nonvanishing-crit}
    m_n(\alpha,\beta) \neq 0 \iff \Omega(\alpha) \cap \Omega(\beta) \neq \varnothing.
\end{equation}
If the intersection is empty, the amplitude vanishes by \eqref{eq:biadj-tree}. Conversely, suppose that $T \in \Omega(\alpha)\cap \Omega(\beta)$. Each internal edge of $T$ defines a factorization channel compatible with both orderings. Taking residues successively in all $n-3$ internal channels of a common cubic tree $T$ isolates the contribution of $T$ since a labeled cubic tree is uniquely determined by its internal splits. The resulting residue is $\pm 1$ and hence $m_n(\alpha,\beta)\neq 0$.

Throughout the paper, nonvanishing is understood as nonvanishing identically on the formal massless, momentum-conserving kinematic space.


In the next section we show how to strengthen the asymptotic upper bound on $\Delta_n$ by using tripartitions.

\section{An Upper Bound from Tripartitions}

The rank of $M_n$ gives the immediate bound $(n-3)!$. To improve this bound, we use the fact that the size of a non-singular diagonal submatrix only depends on which entries of $M_n$ vanish. Let us consider another matrix $K_n$ with the same row and column labels such that
\begin{equation}
    K_n(\alpha,\beta)=0 \iff M_n(\alpha,\beta)=0.
\end{equation}
Any non-singular diagonal submatrix of $M_n$ then maps to a non-singular diagonal submatrix of $K_n$. Consequently,
\begin{equation}
    \Delta_n \leq \rank K_n.
\end{equation}
We are therefore free to construct $K_n$ using auxiliary variables, with its rank computed over their field of rational functions. 

The construction uses the cubic trees contributing to the biadjoint amplitudes. Cutting such a tree at an internal vertex produces three branches, and the vertex can always be chosen so that each branch contains at most $n/2$ labels. We first show that these balanced tripartitions characterize nonzero amplitudes in terms of three smaller nonzero amplitudes. We then construct $K_n$ as a sum of products of these smaller amplitudes. Each tripartition contributes a matrix with a Kronecker-product structure, allowing its rank to be bounded by a product of smaller ranks. The construction can be upgraded to a recursive construction, yielding an exponential improvement over the CHY bound.

\subsection{Tripartitions and Nonvanishing}
By \eqref{eq:nonvanishing-crit}, $m_n(\alpha,\beta)$ is nonzero precisely when there is a cubic tree planar with respect to both orderings $\alpha$ and $\beta$. Cutting such a tree at an internal vertex produces three branches. We will show that the vertex can always be chosen so that each branch contains at most half the external labels. This gives a criterion for nonvanishing in terms of three lower-point double-partial amplitudes.

Let ${\cal O}(E)$ denote the dihedral orderings of a finite label set $E$, and write ${\cal O}_n  = {\cal O}([n])$, where $[n]=\{1,\dots,n\}$. Cutting a cubic tree at an internal vertex $v$ partitions its leaves into three nonempty sets $A$, $B$, and $C$. We call the unordered tripartition $P=\{A,B,C\}$ \textit{balanced} if
\begin{equation}
    |A|, |B|, |C| \leq \frac{n}{2},
\end{equation}
and denote the set of balanced tripartitions by $B_n$. In any planar embedding of the tree, the leaves of each branch occur consecutively along the boundary. For any tripartition $P$, let $I_P \subseteq {\cal O}_n$ be the set of orderings in which each block occurs as a consecutive interval. Introduce an attachment label $\star_X$ for each block $X \in P$, and write $X^\star = X \sqcup \{\star_X\}$. If $\gamma \in I_P$, contracting all labels outside X to $\star_X$ defines the induced local ordering. Explicitly, 
\begin{equation}
    \gamma= (x_1,\dots,x_k,y_1,\dots,y_{n-k}) \implies \gamma\vert_X = [(x_1,\dots,x_k,\star_X)] \in {\cal O}(X^\star),
\end{equation}
where $X= \{x_1,\dots,x_k\}$ and brackets denote the dihedral class. Reflection can change the sign of a double-partial amplitude, but not its vanishing, so the local nonvanishing conditions are independent of the chosen representatives. For a singleton $X=\{x\}$, the local tree is the single edge joining $x$ to $\star_X$. We use the unique two-label ordering and assign its factor the value $m_2 = 1$. This is a convention, not a two-point scattering amplitude.

Each local amplitude is evaluated on its own generic massless, momentum-conserving kinematic space. Indeed, an attachment momentum inherited from the original diagram would satisfy
\begin{equation}
    p_{\star_X} = -\sum_{i \in X}p_i, \quad p_{\star_X}^2 = s_X,
\end{equation}
which is generically nonzero. Ordinary on-shell factorization requires the appropriate pole. Here the cut is only used to determine whether compatible cubic trees exist, so the lower-point amplitudes may instead be assigned independent massless kinematics.

The required balanced vertex exists in every cubic tree with $n\geq 3$ leaves. To see this, start at any internal vertex. If one branch contains $r > n/2$ leaves, move across the edge into that branch. The new vertex is internal, and the branch leading back contains $n-r<n/2$ leaves. The walk therefore never backtracks and, since the tree is finite, must terminate at a vertex with at most $n/2$ leaves in each branch. Consequently, every local amplitude obtained from this vertex has
\begin{equation}
    |X|+1 \leq \lfloor n/2\rfloor+1 < n, \quad n\geq4.
\end{equation}
This reduction in size allows the rank construction.

We can now characterize nonvanishing in terms of balanced tripartitions. 
For $n\geq 3$ and $\alpha,\beta \in {\cal O}_n$, $m_n(\alpha,\beta) \neq 0$ if and only if there exists a balanced tripartition $P \in B_n$ such that $\alpha,\beta \in I_P$ and \begin{equation}
\label{eq:balanced-trip}
    m_{|X|+1}(\alpha\vert_X,\beta\vert_X) \neq 0, \quad \text{for every } X\in P.
\end{equation}
The local amplitudes use the independent kinematics specified above, with $m_2=1$ for singleton blocks.

To establish the criterion, first suppose that $m_n(\alpha,\beta)\neq0$. By \eqref{eq:nonvanishing-crit}, choose a common tree $T\in \Omega(\alpha)\cap \Omega(\beta)$ and cut at a balanced vertex $v$. Each branch is consecutive in both boundary orderings, so the resulting tripartition satisfies $P \in B_n$ and $\alpha,\beta \in I_P$. Adding attachment leaves gives local trees planar with respect to both induced orderings. Thus every local factor is nonzero, by \eqref{eq:nonvanishing-crit}, or by the singleton convention.

Conversely, the local nonvanishing conditions supply three common local trees, using the single edge for a singleton block. Glue their attachment leaves together to form an internal vertex $v$. For each $\gamma \in \{\alpha,\beta\}$, choose embeddings with boundary orderings $\gamma\vert_X$ and arrange them in disjoint sectors around $v$, following the block order in $\gamma$. The glued abstract tree is therefore planar for both orderings, with possibly different embeddings. \eqref{eq:nonvanishing-crit} gives $m_n(\alpha,\beta)\neq0$.
For any fixed tripartition $P$ with $\alpha,\beta \in I_P$, cutting and gluing are inverse operations and give the bijection 
\begin{equation}
    (T,v) \longleftrightarrow (T_A^\star,T_B^\star,T_C^\star),
\end{equation}
where $T \in \Omega(\alpha)\cap\Omega(\beta)$, the internal vertex $v$ has branch sets $A,B,C$, and $T_{X}^\star \in \Omega(\alpha\vert_X) \cap \Omega(\beta\vert_X)$ for every $X \in P$. This bijection holds without balance.

\begin{example}
\label{ex:tripartition}
    \begin{equation*}
        P = \{ A,B,C\}, \qquad A= \{1,2,3\},\quad B=\{4,5,6\},\quad C=\{7\},
    \end{equation*} 
    \begin{equation*}
        \alpha = (1,2,3,4,5,6,7), \qquad\beta = (1,3,2,4,6,5,7).
    \end{equation*}
    The block sizes are $(3,3,1)$, so $P$ is balanced, and each block is consecutive in both orderings. The induced local orderings are 
    \begin{equation}
    \begin{aligned}
        \alpha\vert_A &= [(1,2,3,\star_A)],  &\beta\vert_A &= [(1,3,2,\star_A)] \\
        \alpha\vert_B &= [(4,5,6,\star_B)], &\beta\vert_B &= [(4,6,5,\star_B)]\\
        \alpha\vert_C &= [(7,\star_C)], &\beta\vert_C &= [(7,\star_C)].
    \end{aligned}
    \end{equation}
    The two orderings on $A^\star$ admit a common four-leaf tree whose internal edge separates $\{2,3\}$ from $\{1,\star_A\}$. Similarly, the orderings on $B^\star$ admit the tree separating $\{5,6\}$ from $\{4,\star_B\}$. The local tree on $C^\star$ is the single edge joining 7 to $\star_C$. Identifying the three attachment leaves produces a cubic tree planar with respect to both $\alpha$ and $\beta$.
\end{example}

\begin{figure}[htbp]
  \centering
  \resizebox{\linewidth}{!}{
\begingroup
\definecolor{treeblue}{HTML}{2676BD}
\definecolor{treered}{HTML}{CE4242}
\definecolor{treegreen}{HTML}{278655}
\definecolor{ink}{HTML}{24272B}
\definecolor{muted}{HTML}{6B7278}
\begin{tikzpicture}[
  x=1cm, y=1cm, font=\small, text=ink,
  line width=.85pt, line cap=round, line join=round,
  every node/.style={inner sep=1.5pt},
  vertex/.style={circle,fill=ink,draw=none,inner sep=0pt,minimum size=2.8pt},
  attachment/.style={circle,draw,fill=white,inner sep=0pt,minimum size=5.5pt},
  arrow/.style={-{Stealth[length=1.7mm,width=1.15mm]},draw=muted,line width=.55pt}
]

\foreach \side/\col/\one/\two/\three in
  {-1/treeblue/1/2/3, 1/treered/4/5/6} {
  \begin{scope}[xscale=\side,draw=\col,text=\col]
    \coordinate (u) at (-15:1);
    \path (u) ++(45:1)   coordinate (leaf1)
          (u) ++(-75:1)  coordinate (w)
          (w) ++(-15:1)  coordinate (leaf2)
          (w) ++(-135:1) coordinate (leaf3);
    \draw (0,0)--(u)--(leaf1) (u)--(w)--(leaf2) (w)--(leaf3);
    \node[vertex] at (u) {}; \node[vertex] at (w) {};
    \node[xshift={\side*4pt},yshift=6pt] at (leaf1) {$\one$};
    \node[xshift={\side*7pt},yshift=-2pt] at (leaf2) {$\two$};
    \node[below=4pt] at (leaf3) {$\three$};
  \end{scope}
}
\node at (0,1.77) {$(T,v)$};
\draw[treegreen] (0,0)--(0,1);
\node[text=treegreen,above=4pt] at (0,1) {$7$};
\fill[white] (0,0) circle (3.3pt);
\fill[ink] (0,0) circle (2.5pt);
\node[below=5pt] at (0,0) {$v$};

\draw[arrow] (2.72,-.31)--(3.98,-.31)
  node[midway,above=4pt,font=\footnotesize] {cut at $v$};
\draw[arrow] (3.98,-.65)--(2.72,-.65)
  node[midway,below=4pt,font=\footnotesize] {glue};

\pgfmathsetmacro{\h}{sqrt(3)/2}
\foreach \x/\side/\col/\block/\one/\two/\three in
  {5.7/1/treeblue/A/1/2/3, 8.7/-1/treered/B/4/5/6} {
  \node at (\x,1.77) {$T_{\block^\star}$};
  \node[font=\footnotesize,text=\col] at (\x,1.32)
    {$\block=\{\one,\two,\three\}$};
  \begin{scope}[shift={(\x,-.47)},xscale=\side,draw=\col,text=\col]
    \coordinate (u) at (.5,0);
    \coordinate (w) at (-.5,0);
    \draw (1,\h)--(u)--(w)--(-1,\h)
          (u)--(1,-\h) (w)--(-1,-\h);
    \node[vertex] at (u) {}; \node[vertex] at (w) {};
    \node[above=4pt] at (1,\h) {$\one$};
    \node[above=4pt] at (-1,\h) {$\two$};
    \node[below=4pt] at (-1,-\h) {$\three$};
    \node[attachment] at (1,-\h) {};
    \node[below=5pt] at (1,-\h) {$\star_{\block}$};
  \end{scope}
}

\node at (11.7,1.77) {$T_{C^\star}$};
\node[font=\footnotesize,text=treegreen] at (11.7,1.32) {$C=\{7\}$};
\begin{scope}[shift={(11.7,-.47)},draw=treegreen,text=treegreen]
  \draw (0,-.5)--(0,.5);
  \node[above=4pt] at (0,.5) {$7$};
  \node[attachment] at (0,-.5) {};
  \node[below=5pt] at (0,-.5) {$\star_C$};
\end{scope}
\end{tikzpicture}
\endgroup}
  \caption{The cut-and-glue construction for Example \eqref{ex:tripartition}. Hollow circles denote attachment leaves; identifying them recovers the marked vertex $v$.}
  \label{fig:cut-glue}
\end{figure}
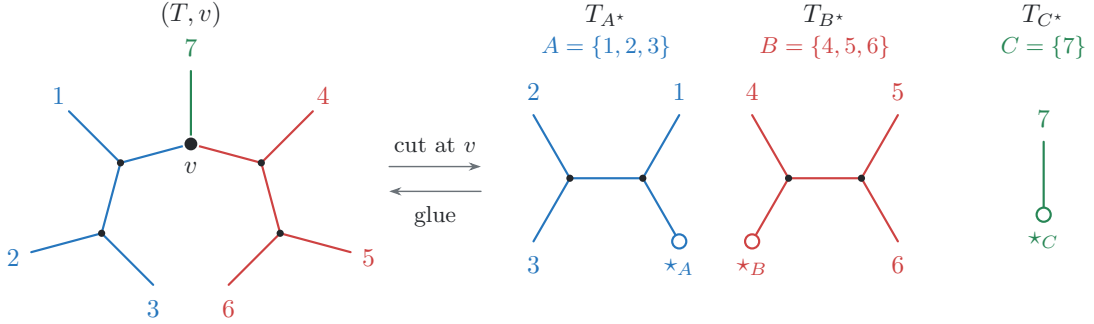
Next we will construct the matrix $K_n$ and show that it has the same zero pattern as $M_n$.   
\subsection{Constructing $K_n$}

We now use the tripartition criterion of \eqref{eq:balanced-trip} to construct a matrix $K_n$ with the same zero pattern as $M_n$, but whose rank can be bounded in terms of lower-point amplitude matrices.

Fix $P=\{A,B,C\}$ and choose the block order $A,B,C$. Every $\alpha \in I_P$ has a unique representative by concatenation,
\begin{equation}
    \alpha = (\sigma_A,\sigma_B,\sigma_C), \quad \sigma_X \in \frak S_X, \quad X \in P,
\end{equation}
where $\frak S_X$ denotes permutations of $X$ \footnote{Note that $S_n = {\frak S}_{[n]}$. In this section we find it useful to explicitly introduce the set being permuted in the notation.}. Indeed, reflect if necessary so the cyclic block order is $A,B,C$, then rotate to the beginning of $A$. Thus,
\begin{equation}
    I_P \cong \frak S_A \times \frak S_B \times \frak S_C.
\end{equation}
In particular, the three permutations can be chosen independently. We use the analogous notation $\beta = (\rho_A,\rho_B,\rho_C)$. Recall that for these linear orderings, we have $\alpha\vert_X = [\sigma_X,\star_X]$.


For each $P \in B_n$, define the matrix $K_n^P$, with rows and columns indexed by ${\cal O}_n$, by
\begin{equation}
    K_n^P(\alpha,\beta) := \begin{cases}
        \displaystyle\prod_{X\in P} m_{|X|+1}\left((\sigma_X,\star_X),(\rho_X,\star_X)\right),  &\alpha,\beta \in I_P \\
        0, &\text{otherwise}.
    \end{cases}
\end{equation}
Here $\alpha=(\sigma_A,\sigma_B,\sigma_C)$ and $\beta = (\rho_A,\rho_B,\rho_C)$ are the unique representatives. The local amplitudes use independent generic massless, momentum-conserving kinematics, with $m_2=1$ for singleton blocks.

The advantage of this definition is that $K_n^P$ has a Kronecker-product structure. To see this, for each $X \in P$, introduce the permutation-indexed local matrix:
\begin{equation}
    \widehat M_X(\sigma,\rho) := m_{|X|+1}((\sigma,\star_X),(\rho,\star_X)), \quad \sigma, \rho \in \frak S_X.
\end{equation}
Since $I_P$ is indexed by triples of permutations, the row and column indices of $\displaystyle{K_n^P}$ factor into three independent components. Moreover, by construction, for $\alpha,\beta \in I_P$, 
\begin{equation}
    K_n^P(\alpha,\beta) = \widehat M_A(\sigma_A,\rho_A)\widehat M_B(\sigma_B,\rho_B)\widehat M_C(\sigma_C,\rho_C).
\end{equation}
This is exactly the entrywise definition of the Kronecker product. Thus, with rows and columns indexed by triples of permutations, the restriction of $\displaystyle{K_n^P}$ to $I_P\times I_P$ is
\begin{equation}
    \left. K_n^P\right|_{I_P\times I_P} = \widehat M_A \otimes \widehat M_B \otimes \widehat M_C.
\end{equation}
All other rows and columns of $K_n^P$ outside $I_P$ vanish. Consequently,
\begin{equation}
    \rank K_n^P = \prod_{X \in P}  \rank \widehat M_X.
\end{equation}
To evaluate these ranks, note that every dihedral ordering of $X^\star$ is represented by some $(\sigma,\star_X)$. Different permutations representing the same dihedral ordering give rows and columns that agree up to reflection sign. Hence $\widehat M_X$ is obtained from $M_{|X|+1}$ by repeating rows and columns up to signs\footnote{Conversely, selecting one permutation for each dihedral ordering would recover $M_{|X|+1}$ as a submatrix. Equivalently, after grouping permutations into reversal pairs, $\widehat M_X$ can be written as $M_{|X|+1} \otimes (\begin{smallmatrix}
    1&\epsilon \\ \epsilon&1 \end{smallmatrix})$, where $\epsilon = (-1)^{|X|+1}$. The $2 \times 2$ factor has rank one, and its support is $J_2 = (\begin{smallmatrix}
        1&1 \\ 1&1\end{smallmatrix})$, which is used in Section 5.}
\begin{equation}
    \rank \widehat M_X =\rank M_{|X|+1}.
\end{equation}
Using the CHY rank bound, we conclude that
\begin{equation}
    \rank K_n^P = \prod_{X \in P}\rank M_{|X|+1} \leq \prod_{X \in P}((|X|+1)-3)! = \prod_{X \in P} (|X|-2)!,
\end{equation}
where we adopt $(-1)!=1$.
It remains to combine the contributions of all the balanced tripartitions while preserving their nonvanishing information. Introduce algebraically independent variables $t_P$, and define
\begin{equation}
    K_n := \sum_{P \in B_n}t_P K_n^P.
\end{equation}
The variables $t_P$ prevent cancellation between contributions from distinct tripartitions. Thus, $K_n(\alpha,\beta)$ vanishes identically if and only if every product in this sum vanishes. More precisely,
\begin{equation}
\begin{aligned}
    K_n(\alpha,\beta)\neq 0 \iff  &\exists P \in B_n \text{ such that } \alpha,\beta \in I_P, \\ &m_{|X|+1}(\alpha\vert_X,\beta\vert_X) \neq 0 \text{ for every } X\in P.
    \end{aligned}
\end{equation}
The tripartition criterion from above therefore gives 
\begin{equation}
    K_n(\alpha,\beta) =0 \iff m_n(\alpha,\beta)=0.
\end{equation}
Here every non-singular diagonal submatrix of $M_n$ therefore selects a non-singular diagonal submatrix of $K_n$. Together with subadditivity of rank, this gives
\begin{equation}
\label{eq:subadditivity-KPn}
    \Delta_n \leq \rank K_n \leq \sum_{P\in B_n} \rank K_n^P \leq \sum_{P\in B_n} \prod_{X \in P}(|X|-2)!.
\end{equation}
We introduce the following notation for convenience as it shows up in Section 4.4:
\begin{equation}
\label{eq:upper-bound}
    U_n := \sum_{P\in B_n} \prod_{X \in P}(|X|-2)!.
\end{equation}

\subsection{Recursive Upper Bound}
The construction of Section 4.2 can be iterated by replacing each local amplitude with a matrix with the same zero pattern to give the upper bound $R_n$. The support argument remains the same, while the local rank bounds are supplied by induction. Set $R_2=R_3=1$ and, for $n\geq 4$, define
\begin{equation}
\label{eq:recursive-bound}
    R_n := \min \bigl\{ (n-3)!, \sum_{P\in  B_n} \prod_{X \in P} R_{|X|+1}\bigr\}.
\end{equation}
This recurrence only involves smaller sizes. Balancedness enforces $2\leq|X|+1 \leq \lfloor n/2\rfloor +1 <n$ for every $X\in P$ and $n\geq 4$. We prove by induction that, for every $n\geq 2$, there is a matrix $W_n$, with rows and columns indexed by ${\cal O}_n$, such that 
\begin{equation}
\label{eq:recursive-statement}
    W_n(\alpha,\beta) = 0 \iff m_n(\alpha,\beta) =0, \qquad \rank W_n \leq R_n.
\end{equation}
As in Section 4.2, vanishing and rank are understood over the relevant rational function fields, rather than at a particular specialization of the variables.

For $n=2,3$, take $W_n=(1)$, using the convention $m_2=1$. Both claims hold trivially. Now fix $n\geq 4$, and assume that both claims hold for every $2\leq k <n$. For each block $X$, take a relabeled copy $W_X$ of $W_{|X|+1}$, indexed by dihedral orderings of $X^\star$. Use independent copies of all local kinematics and auxiliary variables. This preserves nonzero entries and ranks. For each $P \in B_n$, define a matrix $\widetilde W_n^P$ with rows and columns indexed by $\mathcal O_n$,
\begin{equation}
    \widetilde{W}_n^P(\alpha,\beta) := \begin{cases}
        \displaystyle\prod_{X\in P} W_X(\alpha\vert_X,\beta\vert_X), &\alpha,\beta \in I_P \\
        0, &\text{otherwise}.
    \end{cases}
\end{equation}
Introduce $t_P$, algebraically independent over all local coefficient fields, and define
\begin{equation}
    \widetilde{W}_n : = \sum_{P \in B_n} t_P \widetilde{W}_n^P.
\end{equation}
We use subadditivity to obtain
\begin{equation}
    \rank \widetilde W_n \leq \sum_{P \in B_n} \rank \widetilde W_n^P.
\end{equation}
To bound the rank of $\widetilde W_n^P$, fix $P=\{A,B,C\}$ and index $I_P$ by triples of permutations as in Section 4.2. For each $X \in P$, introduce the local matrix
\begin{equation}
    \widehat W_X(\sigma,\rho) := W_X([\sigma,\star_X],[\rho,\star_X]), \quad \sigma,\rho \in \frak S_X.
\end{equation}
Every dihedral ordering of $X^\star$ is represented by some $(\sigma,\star_X)$. Thus $\widehat W_X$ is obtained from $W_X$ by repeating rows and columns. Conversely, selecting one permutation for each dihedral ordering recovers $W_X$ as a submatrix. Consequently,
\begin{equation}
    \rank \widehat W_X = \rank W_X.
\end{equation}
Using the indexing of $I_P$ from Section 4.2, we have
\begin{equation}
    \left. \widetilde W_n^P\right|_{I_P\times I_P} = \widehat W_A \otimes \widehat W_B \otimes \widehat W_C.
\end{equation}
All rows and columns indexed outside $I_P$ vanish, so 
\begin{equation}
    \rank \widetilde W_n^P = \prod_{X \in P}  \rank \widehat W_X= \prod_{X \in P}  \rank W_X.
\end{equation}
Subadditivity of rank and the induction hypothesis therefore give
\begin{equation}
    \rank \widetilde W_n \leq \sum_{P \in B_n} \prod_{X \in P}  \rank W_X \leq \sum_{P \in B_n} \prod_{X \in P} R_{|X|+1}.
\end{equation}

It remains to show that $\widetilde W_n$ has the same zero pattern as $M_n$. Algebraic independence of the $t_P$ implies that an entry of $\widetilde{W}_n$ is nonzero if and only if at least one of $\widetilde{W}_n^P$ is nonzero. By the induction hypothesis, therefore, 
\begin{equation}
\begin{aligned}
      \widetilde{W}_n(\alpha,\beta) \neq 0 \iff &\text{there exists } P \in B_n \text{ with } \alpha,\beta \in I_P \\ 
      &\text{and } m_{|X|+1}(\alpha\vert_X,\beta\vert_X)\neq 0 \text{ for every }  X \in P.  
\end{aligned}
\end{equation}
The balanced tripartition criterion \eqref{eq:balanced-trip} identifies the right side with $m_n(\alpha,\beta)\neq0$. Thus $\widetilde{W}_n$ has the same support as $M_n$.

To complete the inductive step, recall that $\rank M_n \leq (n-3)!$. Choose $W_n = M_n$ if the first bound in \eqref{eq:recursive-bound} is no larger than the second, and $W_n =\widetilde{W}_n$ otherwise. Both choices have the required support, and 
\begin{equation}
    \rank W_n \leq \min \left\{ (n-3)!, \sum_{P\in B_n}\prod_{X\in P} R_{|X|+1}\right\}.
\end{equation}
This proves both assertions in \eqref{eq:recursive-statement} and completes the induction.

Finally, the same row and column labels that select a non-singular diagonal submatrix of $M_n$ select one of $W_n$. Consequently,
\begin{equation}
    \Delta_n \leq \rank W_n \leq R_n.
\end{equation}
In the next subsection, we analyze the asymptotics of the upper bounds.

\subsection{Asymptotic Behavior}
We now analyze the upper bounds $U_n$ and $R_n$ obtained above. A single application of the tripartition construction gives a suppression of order $\log n/n$ relative to the CHY rank, while the recursive construction gives an exponential suppression. We normalize both bounds by writing
\begin{equation}
    u_n:= \frac{U_n}{(n-3)!}, \qquad r_n:= \frac{R_n}{(n-3)!}, \qquad n\geq 3,
\end{equation}
and set $r_2=1$.

\begin{table}[tbp]
\centering
\small
\setlength{\tabcolsep}{4pt}
\begin{tabular}{@{}rrr@{\qquad}rrr@{\qquad}rrr@{}}
\toprule
$n$ & $u_n$ & $r_n$
& $n$ & $u_n$ & $r_n$
& $n$ & $u_n$ & $r_n$ \\
\cmidrule(r){1-3}
\cmidrule(lr){4-6}
\cmidrule(l){7-9}
52 & 1.392259 & 1.000000
& 71 & 0.772651 & 0.772651
& 99 & 0.574292 & 0.574292 \\

53 & 1.005134 & 1.000000
& 72 & 1.019025 & 1.000000
& 100 & 0.747509 & 0.747509 \\

54 & 1.342334 & 1.000000
& 73 & 0.753670 & 0.753670
& 107 & 0.535944 & 0.535944 \\

55 & 0.972049 & 0.972049
& 74 & 0.992855 & 0.992855
& 108 & 0.695566 & 0.686399 \\

56 & 1.296034 & 1.000000
& 75 & 0.735661 & 0.735661
& 119 & 0.487653 & 0.469862 \\

57 & 0.941220 & 0.941220
& 76 & 0.968060 & 0.968060
& 120 & 0.630439 & 0.580036 \\

58 & 1.252971 & 1.000000
& & &
& 149 & 0.399483 & 0.305190 \\

& &
& & &
& 150 & 0.512417 & 0.409635 \\

& &
& & &
& 200 & 0.392822 & 0.190108 \\
\bottomrule
\end{tabular}
\caption{Selected values of the normalized upper bounds
$u_n=U_n/(n-3)!$ and $r_n=R_n/(n-3)!$,
rounded to six decimal places.
The odd and even subsequences of $u_n$ first fall below
$1$ at $n=55$ and $n=74$, respectively.
The recursive bound first improves strictly on
$\min\{1,u_n\}$ at $n=108$.}
\label{tab:upper-bounds}
\end{table}

Let us start by writing $U_n$ in a form that is more amenable to asymptotics. The original bound was written as a sum over unordered tripartitions. Here we write it as a sum over ordered balanced tripartitions. We denote $a=|A|$, $b=|B|$, $c=|C|$ and use the convention $(-1)!=1$. Thus,
\begin{equation}
    U_n = \frac{1}{3!} \sum_{\substack{a+b+c=n \\ 1\leq a,b,c \leq \lfloor n/2 \rfloor}} {n \choose a,b,c} (a-2)! (b-2)! (c-2)!.
\end{equation}
Expanding out this multinomial coefficient and normalizing by $(n-3)!$, we can write this sum as
\begin{equation}
    u_n = \frac{n(n-1)(n-2)}{6}\sum_{\substack{a+b+c=n \\ 1\leq a,b,c\leq \lfloor n/2\rfloor}} \frac{(a-2)!}{a!} \frac{(b-2)!}{b!} \frac{(c-2)!}{c!}.
\end{equation}
Let $H_k = \sum_{j=1}^kj^{-1}$, and $H_k^{(2)}=\sum_{j=1}^kj^{-2}$, with $H_0=H_0^{(2)}=0$. For $m\geq 3$, we have the exact identities\footnote{These formulas were obtained by AI tools and numerically checked.},
\begin{equation}
\label{eq:even-odd-un}
\begin{aligned}
    u_{2m} &= \left( 4+ \frac{6}{m-1} +\frac{4}{m}\right) \frac{H_{m-3}}{m-2} + \frac{3 H_{m-2}^{(2)}}{2m-3}+ \frac{9}{m-2} +\frac{3}{m-1} + \frac{3}{m}, \\
    u_{2m+1} &= \left( 4+ \frac{3}{m-1} +\frac{1}{m}\right) \frac{H_{m-2}}{m-1} + \frac{3 H_{m-2}^{(2)}}{2m-2}+ \frac{7}{m-1} +\frac{1}{2(m-1)^2}.
    \end{aligned}
\end{equation}
For the asymptotic behavior, inserting 
\begin{equation}
    H_k = \log k + \gamma + O(k^{-1}), \quad H_k^{(2)} = \frac{\pi^2}{6} + O(k^{-1}),
\end{equation}
into the identity gives
\begin{equation}
    u_n = \frac{8 \log n}{n} +O\left(\frac{1}{n}\right), \quad n \to \infty.
\end{equation}


Thus, already at this stage, the diagonal degree occupies a vanishing fraction of the CHY rank. More explicitly, the even and odd subsequences in \eqref{eq:even-odd-un} are decreasing for $m\geq 4$. Since $u_{73} < 377/500$ and $u_{74}<993/1000$, it follows that
\begin{equation}
    \Delta_n \leq R_n \leq U_n <(n-3)!, \quad n\geq 73.
\end{equation}

We now turn to the recursive bound and obtain the exponential bound on $r_n$. Set $\rho=99/100$. We claim that 
\begin{equation}
\label{eq:recursive-estimate}
    r_n\leq\rho^{n-74}, \quad n\geq 2.
\end{equation}
For $n\leq 74$, this follows from $r_n\leq 1$. The exact evaluation of the recurrence for $R_n$ verifies the bound for $75 \leq n \leq 352.$ The harmonic identities give
\begin{equation}
    u_{353}, u_{354} < \frac{93}{400} < \rho^{145}.
\end{equation}
Monotonicity of the even and odd subsequences therefore implies $u_n<\rho^{145}$ for all $n\geq353$.

Now fix $n\geq 353$ and assume the claimed bound at every smaller size. For each balanced tripartition $P$, every block satisfies $|X|+1<n$, so the induction hypothesis applies to all three factors in the recurrence and gives\footnote{Balancedness gives $2\leq |X|+1<n$ so the induction hypothesis \eqref{eq:recursive-estimate} applies at size $|X|+1$. For $|X|+1\leq74$, the estimate follows simply from $r_{|X|+1}\le1\le\rho^{|X|+1-74}$. Although loose for small blocks, this uniform bound makes the product of the three exponential factors equal to $\rho^{n-219}$, independently of the tripartition.} 
\begin{equation}
    R_{|X|+1} = (|X|-2)! r_{|X|+1} \leq (|X|-2)! \rho^{|X|+1-74}.
\end{equation}
Multiplying the three gives the factor $\rho^{n+3-3\cdot 74} = \rho^{n-219}$, since $\sum_{X \in P} |X|=n$. Hence,
\begin{equation}
\begin{aligned}
        R_n &\leq \sum_{P \in B_n} \prod_{X \in P}R_{|X|+1} \\
        &\leq\rho^{n-219} \sum_{P \in B_n} \prod_{X \in P} (|X|-2)! = \rho^{n-219} U_n.
\end{aligned}
\end{equation}
Dividing by $(n-3)!$ gives
\begin{equation}
    r_n \leq \rho^{n-219}u_n \leq \rho^{n-219}\rho^{145} = \rho^{n-74},
\end{equation}
completing the induction. Thus, the diagonal degree is exponentially suppressed relative to the CHY rank,
\begin{equation}
    \frac{\Delta_n}{(n-3)!}\leq \left(\frac{99}{100}\right)^{n-74} = O\left( e^{-cn} \right), \quad c=\log(100/99) > 0.01.
\end{equation}
Hence for $n\geq73$, no choice of $(n-3)!$ row and column orderings produces an invertible diagonal double-partial amplitude matrix.

\section{KLT Block Lower Bound}


In the original KLT paper \cite{KLT}, their first construction leads to a formula with $(n-1)!/2\times (n-1)!/2$ terms. As also discussed in their original work, an appropriate  contour deformation reduces the number of terms down to 

%
\begin{align}
    (n-3)!\left( \frac{n-3}{2} \right)!\left( \frac{n-3}{2} \right)! & \quad {\rm for}~n~{\rm odd}, \\
    (n-3)!\left( \frac{n-4}{2} \right)!\left( \frac{n-2}{2} \right)! & \quad {\rm for}~n~{\rm even}.
\end{align}
These formulas have a more tantalizing form if written in terms of $d:=\left( \frac{n-3}{2} \right)!^2$ (for $n$ odd) as $\binom{n-3}{(n-3)/2}\times d^2$. The reason is that this comes from the fact that the matrix of biadjoint amplitudes used in the construction has a block diagonal form with $\binom{n-3}{(n-3)/2}$ blocks of size $d\times d$.

Let us write the KLT formula that leads to the block diagonal basis in the context of quantum field theory as presented in appendix A of \cite{BernKLT} (for $n$ even),
\begin{equation}
\label{kLT-I}
    {\cal A}^{\rm gravitons}_n = \left[ A_n(1,2,\ldots ,n)\sum_{\rm perms}f(I)\bar{f}(J)A_n(1,I,n,J,n-1)\right] + {\cal P}(2,\ldots ,n-2),
\end{equation}
where ${\cal P}(S)$ denotes permutations of the set $S$ and `$perms$' means a sum over $I\in {\cal P}(2,3,\ldots ,n/2)$ and $J\in {\cal P}(n/2+1,\ldots, n-2)$. Also, $f(I)$ and $\bar{f}(J)$ are polynomials in $s_{ab}$\footnote{While the KLT construction manifestly produces polynomials, this is a very surprising property from the CHY perspective, as these functions arise from the inverse of a matrix of biadjoint scalar amplitudes.}.   

We next exhibit the block diagonal form and make the connection to the matrix of biadjoint scalar amplitudes explicit.

\subsection{KLT Block Construction}
\label{sec:klt-review}

Set
\begin{equation}
 q:=n-3,
 \qquad
 r:=\left\lceil\frac q2\right\rceil,
 \qquad
 s:=\left\lfloor\frac q2\right\rfloor,
 \qquad r+s=q,
 \label{eq:qrs}
\end{equation}
and let
\begin{equation}
 S:=\{2,3,\ldots,n-2\}.
\end{equation}
Thus $|S|=q$. The KLT row and column sets are
\begin{align}
 \cA
 &:=\left\{
 (1,\omega(2),\omega(3),\ldots,\omega(n-2),n-1,n)
 :\omega\in\Sn_{S}
 \right\},
 \label{eq:KLT-A}\\
 \cB
 &:=\left\{
 (1,\gamma(2),\ldots,\gamma(r+1),n,
 \gamma(r+2),\ldots,\gamma(n-2),n-1)
 :\gamma\in\Sn_{S}
 \right\}.
 \label{eq:KLT-B}
\end{align}
Both sets have $(n-3)!=q!$ elements.

The blocks are indexed by partitions of $S = H\cup G$
\footnote{Each partition $H \mid G$ also determines the balanced tripartition $P_H = \left\{A,\{n\},C\right\}, \quad A = \{1\} \cup H, \quad C = G \cup \{n-1\}$. The row and column orderings of the corresponding KLT block belong to $I_{P_H}$, connecting this construction to \eqref{eq:balanced-trip}.} with $|H|=r$, and $|G|=s$.

The two sides are distinguished: $H$ is placed before $n$ in the column ordering and $G$ after $n$. In particular, when $r=s$, the choices $H\mid G$ and $G\mid H$ label different KLT blocks.

For linear orderings $\sigma$ of $H$ and $\tau$ of $G$, write $\sigma(H)$ and $\tau(G)$ for the corresponding lists. Define
\begin{align}
 \cA_H
 &:={\left\{
 (1,\sigma(H),\tau(G),n-1,n)
 :\sigma\in\Sn_H,\ \tau\in\Sn_G
 \right\}},
 \label{eq:AH}\\
 \cB_H
 &:={\left\{
 (1,\sigma'(H),n,\tau'(G),n-1)
 :\sigma'\in\Sn_H,\ \tau'\in\Sn_G
 \right\}}.
 \label{eq:BH}
\end{align}
Each block has
\begin{equation}
 |\cA_H|=|\cB_H|=r!s!=d
 \label{eq:block-size}
\end{equation}
rows and columns. The sets $\cA_H$ partition $\cA$ and the sets $\cB_H$ partition $\cB$.

Now we can rewrite the KLT formula \eqref{kLT-I} in a way that manifests the block structure,
\begin{equation}
    \label{KLT-II}
    \begin{aligned}
    {\cal A}^{\rm gravitons}_n = \sum_{\substack{H \subset S \\ |H|=r}}\sum_{\substack{\sigma,\sigma'\in \Sn_H \\ \tau,\tau' \in \Sn_G}}&A_n(1,\sigma(H),\tau(G),n-1,n) \\
    &\times S[\sigma,\tau \mid \sigma',\tau']A_n(1,\sigma'(H),n,\tau'(G),n-1).
    \end{aligned}
\end{equation}
The choice of $H$ labels a KLT block. The second sum is over all $r!$ permutations $\sigma$ and $\sigma'$ of $H$ and over all $s!$ permutations $\tau$ and $\tau'$ of $G$. The permutations $(\sigma,\tau)$ and $(\sigma',\tau')$ label the rows and columns of the KLT matrix.

Now we are ready to build the matrix of biadjoint amplitudes that gives rise to the block diagonal KLT matrix.

Consider the submatrix of $M_n$, the matrix of biadjoint amplitudes, with rows and columns labeled by $(n-3)!$ orderings organized in blocks labeled by the choice of $H$ and with row and column indices $(\sigma,\tau)$ and $(\sigma',\tau')$ as above. We want to prove that this submatrix is indeed block diagonal.

Consider $H_1\neq H_2$, set $G_i = S \setminus H_i$, and take an entry joining $\cA_{H_1}$ to $\cB_{H_2}$:
\begin{equation}
\label{eq:off-diagonal-block}
    m_n((1,\sigma(H_1),\tau(G_1),n-1,n),(1,\sigma'(H_2),n,\tau'(G_2),n-1)).
\end{equation}
Because $|H_1|=|H_2|$, there are labels $q_1\in H_1 \setminus H_2$ and $q_2 \in H_2 \setminus H_1$. In particular, $q_1 \in G_2$ and $q_2 \in G_1$.
Suppose that a cubic tree were planar with respect to both orderings in \eqref{eq:off-diagonal-block}. Prune all leaves except $\{ 1,q_1,q_2,n-1,n\}$. Pruning preserves planarity with respect to the induced orderings, so the resulting five-leaf tree would be planar with respect to both
\begin{equation}
    (1,q_1,q_2,n-1,n) \quad \text{ and } \quad (1,q_2,n,q_1,n-1).
\end{equation}
Every cubic tree with at least four leaves has a cherry\footnote{Two leaves attached to the same vertex.}, whose two leaves must be adjacent in any compatible cyclic ordering. But no pair of labels is adjacent in both of these five-point orderings. Therefore, no such tree exists, and
\begin{equation}
    \Omega(1,q_1,q_2,n-1,n)\cap \Omega(1,q_2,n,q_1,n-1) = \varnothing.
\end{equation}
By the nonvanishing criterion \eqref{eq:nonvanishing-crit}, the entry \eqref{eq:off-diagonal-block} is zero. Consequently, the submatrix of $M_n$ with row set $\cA$ and column set $\cB$ is block diagonal with blocks indexed by $H \subset S$ with $|H|=r$.

\subsection{Direct KLT Lower Bound}

We have shown that the $(n-3)!\times (n-3)!$ matrix obtained by using the KLT construction is block diagonal. Now, selecting a non-zero matrix element from each block produces a list of rows and a list of columns that gives rise to a submatrix with the support of a diagonal matrix of size equal to $\binom{n-3}{s}\times \binom{n-3}{s}$. Recall that $s=\lfloor \frac{(n-3)}{2}\rfloor$. This provides a very loose lower bound
\begin{equation}
\label{lower_bound_I}
    \Delta_n \geq \binom{n-3}{s}.
\end{equation}
The central binomial coefficient gives
\begin{equation}
 \binom{n-3}{\left\lfloor (n-3)/2\right\rfloor}
 =\Theta\!\left(\frac{2^n}{\sqrt n}\right).
 \label{eq:original-asymptotic}
\end{equation}

\subsection{Recursive KLT Lower Bound}

We can improve the direct KLT lower bound \eqref{lower_bound_I} by noticing the natural recursive structure of the formula for KLT blocks. 

Without loss of generality, consider the canonical block defined by $H=\{2,3,\ldots, (n-1)/2\}$ and $G=\{ (n+1)/2,\ldots, n-2\}$ for odd $n$. Clearly, $|H|=|G|= (n-3)/2$. 

Consider a biadjoint amplitude in this block, i.e., 
\begin{equation}
\label{eq:amp_block}
    m_n((1,\sigma(H),\tau(G),n-1,n),(1,\sigma'(H),n,\tau'(G),n-1)).
\end{equation}
 The only way this can happen is that the trees have a vertex that partitions the set of labels into $\{ 1,2,\ldots ,(n-1)/2\}\cup \{n\} \cup \{ (n+1)/2,\ldots ,n-2,n-1\}$. So, the leaf corresponding to the label $n$ is attached to the special vertex.

It is then clear that the amplitude \eqref{eq:amp_block} can be written as follows,
\begin{equation}
    m_L((1,\sigma(H),\star),(1,\sigma'(H),\star))\frac{1}{s_Ls_R}m_R((\star,\tau(G),n-1),(\star,\tau'(G),n-1)),
\end{equation}
where $\star$ represents an off-shell particle (for the biadjoint scalar, a single off-shell particle does not change its combinatorial structure), while $m_L$ and $m_R$ are biadjoint amplitudes with $(n+1)/2$ particles. The two propagators $1/s_L$ and $1/s_R$ are defined using the sets $L=\{1,\sigma(H)\}$ and $R=\{ (n+1)/2,\ldots ,n-2,n-1 \}$ and hence are universal for all amplitudes within the same KLT block.  

Let us denote the KLT block under consideration $K_{H}$ (with the irrelevant factor $s_Ls_R$ removed). Its matrix elements are given, in multi-index notation, by
\begin{equation}
   K_H((\sigma,\tau), (\sigma',\tau')) = m_L(\sigma,\sigma')m_R(\tau,\tau'), 
\end{equation}
with
\begin{align}
    m_L(\sigma,\sigma'):=&  m_L((1,\sigma(H),\star),(1,\sigma'(H),\star)),\\ m_R(\tau,\tau'):= & m_R((\star,\tau(G),n-1),(\star,\tau'(G),n-1)).
\end{align}
Clearly, $K_H = m_L \otimes m_R$. The appearance of this Kronecker structure might be surprising. However, this structure is already present in the KLT formula \eqref{kLT-I}. Note that both $m_L$ and $m_R$ are invertible. To see this, consider that they both are $\frac{(n-3)}{2}!\times \frac{(n-3)}{2}!$ matrices. Also, their entries are amplitudes on $(n+1)/2$ particles. If they were all on-shell, the rank would be $((n+1)/2-3)! = ((n-5)/2)!$. However, one of the particles is off-shell and this increases the rank to $(((n+1)/2+1)-3)!=((n-3)/2)!$. The object that enters in the KLT formula \eqref{kLT-I} is $(K_H)^{-1} = (m_L)^{-1} \otimes (m_R)^{-1}$, which explains the structure $f(I)\bar f(J)$ (see \cite{Bjerrum-Bohr:2010pnr} for a construction that leads to blocks of different sizes).  

The next step is to realize that in any non-zero amplitude of the form
\begin{equation}
    m_L((1,\sigma(H),\star),(1,\sigma'(H),\star)),
\end{equation}
there is always a tree that has labels $1,\star$ in a cherry. This means that if we replace these amplitudes with amplitudes where $1$ and $\star$ are identified, i.e.,  
\begin{equation}
    m_L((1,\sigma(H),\star),(1,\sigma'(H),\star))\to {\tilde m}_L((1,\sigma(H)),(1,\sigma'(H))),
\end{equation}
then the new matrix would have the same support as the original. So,
\begin{equation}
   {\rm supp}(K_H) = {\rm supp}(m_L\otimes m_R) = {\rm supp}({\tilde m}_L\otimes {\tilde m}_R). 
\end{equation}
The new matrices ${\tilde m}_L$ and ${\tilde m}_R$ have components that repeat due to the dihedral invariance, i.e. $\tilde m_L(1,\sigma(H))= \tilde m_L(1,\sigma_R(H))$, where $\sigma_R$ means reversing the order\footnote{This is true up to a sign that only depends on the number of labels and hence it is not relevant to us.}.  
This means that ${\tilde m}_L$ itself has a tensor product structure, i.e., $\supp{\tilde m}_L = \supp({\tilde m}^{\rm red}_L)\otimes J_2$, where
\begin{equation}
    J_2 = \begin{pmatrix}
  1 & 1 \\
  1 & 1
\end{pmatrix}.
\end{equation}
Now we are ready to conclude the argument. Simply note that ${\tilde m}^{\rm red}_L$ is a scaled down version of the original matrix $M_n$. In other words, it is a matrix of biadjoint amplitudes on $n_{\rm red} := (n-1)/2$ points of size $(n_{\rm red}-1)!/2 \times (n_{\rm red}-1)!/2$. Its pattern of zeros follows the same combinatorial structure in terms of compatible trees. This means that one can find a rank $\Delta_{n_{\rm red}}$ diagonal submatrix in it. 

The properties of tensor products and repeating the same argument for $m_R$ then lead us to the conclusion that it is possible to find a submatrix of $K_H$ whose support is that of a non-singular diagonal matrix of rank $\Delta_{n_{\rm red}}^2$.

Going back to the original KLT matrix we conclude that it has a non-singular diagonal submatrix of at least size $\binom{n-3}{(n-3)/2}\Delta_{n_{\rm red}}^2$.

Using that the original KLT matrix is a submatrix of the $M_n$ matrix we arrive at our recursive equation to compute the lower bound of the diagonal degree,
\begin{equation}
    \Delta_n \geq \binom{n-3}{(n-3)/2}\Delta_{(n-1)/2}^2.
\end{equation}

Repeating the same argument for $n$ even is straightforward but the notation is more cumbersome; we leave it as an exercise for the reader.

Combining the odd and even cases, the relation for all $n$ can be written as follows. For $n\geq7$, let
\begin{equation}
 r=\left\lceil\frac{n-3}{2}\right\rceil,
 \qquad
 s=\left\lfloor\frac{n-3}{2}\right\rfloor.
\end{equation}
Then
\begin{equation}
 \DeltaDiag_n\geq
 \binom{n-3}{r}\,
 \DeltaDiag_{r+1}\,
 \DeltaDiag_{s+1}.
 \label{eq:main-recursion}
\end{equation}

\subsection{Asymptotic Behavior of the Recursive KLT Lower Bound}

The relation \eqref{eq:main-recursion} can be used to produce a lower bound for the diagonal degree by finding the asymptotic behavior of the recursion
\begin{equation}
 L_n
 :=\binom{n-3}{r}L_{r+1}L_{s+1},
 \qquad
 r=\left\lceil\frac{n-3}{2}\right\rceil,
 \quad
 s=\left\lfloor\frac{n-3}{2}\right\rfloor,
 \label{eq:L-recursion}
\end{equation}
with initial conditions
\begin{equation}
L_3=1,\quad 
 L_4=1,
 \quad
 L_5=2,
 \quad
 L_6=3,\quad L_7=6.
 \label{eq:seeds}
\end{equation}
Using these conditions and the recursion \eqref{eq:L-recursion} for $n=5,6$, one finds that $L_2^2=1$, $L_2L_3=1$, which we solve by setting $L_2=1$.
Let us introduce
\begin{equation}
 D_n:=\frac{(n-3)!}{L_n}.
 \label{eq:R-def}
\end{equation}
Since $r+s=n-3$, for $n\geq7$, equation \eqref{eq:L-recursion} gives
\begin{align}
 D_n
 &=\frac{(n-3)!}
 {\binom{n-3}{r}L_{r+1}L_{s+1}}
 =\frac{r!s!}{L_{r+1}L_{s+1}}
 \notag\\
 &=r(r-1)s(s-1)D_{r+1}D_{s+1}.
 \label{eq:R-recursion}
\end{align}
Taking logarithms,
\begin{equation}
 \log D_n
 =\log D_{r+1}+\log D_{s+1}
 +\log\bigl(r(r-1)s(s-1)\bigr).
 \label{eq:logR-recursion}
\end{equation}
The final term is $O(\log n)$, while both children have size $n/2+O(1)$. A standard recursion-tree estimate gives
\begin{equation}
 \log D_n=O(n).
 \label{eq:logR-On}
\end{equation}

It follows that there is a constant $C>1$, depending only on the finite set of seed bounds, such that
\begin{equation}
 \DeltaDiag_n\geq L_n\geq\frac{(n-3)!}{C^n}.
 \label{eq:factorial-bound-C}
\end{equation}
Equivalently,
\begin{equation}
 \DeltaDiag_n\geq (n-3)!e^{-O(n)}.
 \label{eq:factorial-bound}
\end{equation}

In appendix A, we make the exponential bound explicit by showing that one may take $C=e^{3/2}$ in \eqref{eq:factorial-bound-C}, yielding
\begin{equation}
    \frac{\Delta_n}{(n-3)!} \geq e^{-3n/2}.
\end{equation}

\section{Discussion}

In this work we have proven that the diagonal degree of the matrix of biadjoint amplitudes $M_n$ is exponentially suppressed relative to the CHY rank, in the large $n$ limit, as follows 
\begin{equation}
    \frac{\Delta_n}{(n-3)!} =  \exp(-\Theta(n)).
\end{equation}
More precisely, we establish the bounds
\begin{equation}
    \exp(-1.5n)\leq\frac{\Delta_n}{(n-3)!}\leq \exp \left( -0.01n+O(1)\right).
\end{equation}
Hence, no diagonal submatrix can attain the full rank $(n-3)!$ for sufficiently large $n$.

This upper bound is the most significant result as it implies that it is not possible to build amplitudes of gravitons as a sum over only $(n-3)!$ products of two gluon partial amplitudes when $n$ is large. 

It is remarkable that a good lower bound is easily obtained by using the original construction proposed in the 1986 KLT work \cite{KLT}. Of course the upper and lower bounds are still far apart and it would be desirable to close this gap.

It is worth mentioning that the function constructed using tripartitions to upper bound the diagonal degree starts with values larger than $(n-3)!$ and it only crosses the $(n-3)!$ line around $n =55$, when $n$ is odd and $n=74$ when $n$ is even. This means that it is still possible that the $(n-3)!$ CHY upper bound can be attained for intermediate values $7\leq n \leq 54$.

Preliminary results using GPT-6 Astra seem to indicate that similar arguments to the ones used in this work can be refined to produce 
\begin{equation}
     \exp( -0.512n)\leq \frac{\Delta_n}{(n-3)!} \leq \exp(-0.025n).
\end{equation}
Our aim here is to give a pedagogical account of the arguments establishing exponential suppression of the diagonal degree relative to the CHY bound. We therefore do not pursue these possible refinements of the exponential constants.

This work was mainly written from a physics perspective. Let us now discuss what our results imply from a mathematics perspective. 

In \cite{FCQ}, the diagonal degree problem was framed in the context of the tropical Grassmannian, ${\rm Trop}\, G(2,n)$, introduced by Speyer and Sturmfels in \cite{SSTrop}, who also showed that it is the moduli space of phylogenetic trees studied by Billera, Holmes and Vogtmann (BHV) \cite{BHV}. Motivated by the work of Postnikov \cite{Alex} on totally positive Grassmannians, $G^+(k,n)$, Speyer and Williams introduced positive tropical Grassmannians  ${\rm Trop}^+\, G(k,n)$ \cite{SWTrop}. Clearly, ${\rm Trop}^+\, G(2,n)$ is the moduli space of planar trees. For a given ordering of $n$ points, $\alpha$, one can define ${\rm Trop}^\alpha\, G(2,n)$ as the subspace of ${\rm Trop}\, G(2,n)$ composed of trees planar with respect to $\alpha$. So, the standard ${\rm Trop}^+\, G(2,n)$ can be thought of as the space of trees planar with respect to the canonical ordering $(1,2,\ldots, n)$. 

In \cite{FCQ}, an intersection matrix $I_n$ with entries $(I_n)_{\alpha, \beta}$ was defined to be $1$ if ${\rm Trop}^\alpha\, G(2,n)$ and ${\rm Trop}^\beta\, G(2,n)$ share at least one maximal cone, i.e., a trivalent tree, and $0$ otherwise. Clearly the $(n-1)!/2\times (n-1)!/2$ matrix $I_n$ has the same support as $M_n$ and the notion of diagonal degree carries over unchanged. 

From the tropical Grassmannian viewpoint, the CHY bound $\Delta_n \leq (n-3)!$ is already quite nontrivial. In fact, there is no known purely combinatorial proof of this. In contrast, the KLT lower bound presented in this work is purely combinatorial. It would be interesting to find a combinatorial proof of the upper bound we have found.

\section*{Acknowledgements}

The authors thank M. Froese, K. Shahini, and B. Umbert for useful discussions, and R. Yeats for useful discussions and valuable comments on the draft that helped improve the manuscript. We also acknowledge the use of ChatGPT for the initial explorations that gave rise to the balanced tripartition construction of the upper bound. We believe that this was motivated by the extreme balanced tripartition already used in the KLT construction given as part of the input. ChatGPT and Claude were also used to help us streamline the arguments for the asymptotic behavior of the various recursion relations. Research at Perimeter Institute is supported in part by the Government of Canada through the Department of Innovation, Science and Economic Development Canada and by the Province of Ontario through the Ministry of Colleges and Universities.

\appendix

\section{An Explicit $C$ for the Lower Bound}
Set $N=n+1$ and $F_N=D_{N-1}$. If $a=\lceil N/2\rceil$ and $b = \lfloor N/2\rfloor$, then $a+b =N$ and the recursion becomes
\begin{equation}
    F_N = (a-2)(a-3)(b-2)(b-3)F_aF_b.
\end{equation}
We claim that for $N\geq 6$
\begin{equation}
    F_N\leq \frac{e^{3N/2}}{16N^2(N-2)(N-3)}.
\end{equation}
We obtain the following base cases from \eqref{eq:seeds}:
\begin{equation}
    F_6=1,\quad F_7 = 2,\quad F_8=4,\quad F_9=12,\quad F_{10}=36,\quad F_{11}=72.
\end{equation}
Direct substitutions verify the claim for $6\leq N\leq 11$.

We now proceed with the inductive step. For $N\geq 12$, both $a,b\geq 6$. Applying the inductive hypothesis gives 
\begin{equation}
\begin{aligned}
    F_N &\leq (a-2)(a-3)(b-2)(b-3) \frac{e^{3a/2}}{16a^2(a-2)(a-3)} \frac{e^{3b/2}}{16b^2(b-2)(b-3)} \\
    &= \frac{e^{3N/2}}{256a^2b^2}.
    \end{aligned}
\end{equation}
Now use $ab = \lfloor N^2/4\rfloor \geq \frac{N^2-1}{4}$. Consequently, $16a^2b^2\geq (N^2-1)^2 \geq N^2(N-2)(N-3)$. Therefore we close the induction with
\begin{equation}
    F_N\leq \frac{e^{3N/2}}{16N^2(N-2)(N-3)}.
\end{equation}
Since $N=n+1$, we have
\begin{equation}
    D_n\leq \frac{e^{3(n+1)/2}}{16(n+1)^2(n-1)(n-2)}.
\end{equation}
For $n\geq 5$, the denominator is greater than $e^{3/2}$. Hence,
\begin{equation}
    D_n \leq e^{3n/2}.
\end{equation}
Recalling $\Delta_n \geq L_n = \frac{(n-3)!}{D_n}$, we conclude
\begin{equation}
    \frac{\Delta_n}{(n-3)!}\geq e^{-3n/2}.
\end{equation}

\bibliographystyle{JHEP}
\bibliography{references}

@article{Dolan:2013isa,
    author = "Dolan, Louise and Goddard, Peter",
    title = "{Proof of the Formula of Cachazo, He and Yuan for Yang-Mills Tree Amplitudes in Arbitrary Dimension}",
    eprint = "1311.5200",
    archivePrefix = "arXiv",
    primaryClass = "hep-th",
    doi = "10.1007/JHEP05(2014)010",
    journal = "JHEP",
    volume = "05",
    pages = "010",
    year = "2014"
}

@article{Bjerrum-Bohr:2010pnr,
    author = "Bjerrum-Bohr, N. E. J. and Damgaard, Poul H. and Sondergaard, Thomas and Vanhove, Pierre",
    title = "{The Momentum Kernel of Gauge and Gravity Theories}",
    eprint = "1010.3933",
    archivePrefix = "arXiv",
    primaryClass = "hep-th",
    reportNumber = "IPHT-T10-14, IHES-P-10-45",
    doi = "10.1007/JHEP01(2011)001",
    journal = "JHEP",
    volume = "01",
    pages = "001",
    year = "2011"
}

@article{FCQ,
    author = "Cachazo, Freddy",
    title = "{Diagonally Embedded Sets of ${\rm Trop}^+G(2,n)$'s in ${\rm Trop}\, G(2,n)$: Is There a Critical Value of $n$?}",
    eprint = "2104.10628",
    archivePrefix = "arXiv",
    primaryClass = "math.CO",
    month = "4",
    year = "2021"
}

@article{BHV,
 author = {Billera, Louis J. and Holmes, Susan P. and Vogtmann, Karen},
 title = {Geometry of the Space of Phylogenetic Trees},
 journal = {Adv. Appl. Math.},
 issue_date = {November 2001},
 volume = {27},
 number = {4},
 month = nov,
 year = {2001},
 issn = {0196-8858},
 pages = {733--767},
 numpages = {35},
 url = {http://dx.doi.org/10.1006/aama.2001.0759},
 doi = {10.1006/aama.2001.0759},
 acmid = {510852},
 publisher = {Academic Press, Inc.},
 address = {Orlando, FL, USA},
}

@ARTICLE{SWTrop,
       author = {{Speyer}, David and {Williams}, Lauren K.},
        title = "{The tropical totally positive Grassmannian}",
      journal = {arXiv Mathematics e-prints},
         year = 2003,
        month = dec,
          eid = {math/0312297},
        pages = {math/0312297},
archivePrefix = {arXiv},
       eprint = {math/0312297},
 primaryClass = {math.CO},
       adsurl = {https://ui.adsabs.harvard.edu/abs/2003math.....12297S}
}

@article{SSTrop,
  title="{The tropical Grassmannian}",
  author={Speyer, David and Sturmfels, Bernd},
  journal={Advances in Geometry},
  volume={4},
  number={3},
  pages={389--411},
  year={2004},
  publisher={Berlin; New York: Walter de Gruyter},
  eprint         = "math/0304218",
      archivePrefix  = "arXiv",
      primaryClass   = "math",
}

@ARTICLE{Alex,
       author = {{Postnikov}, Alexander},
        title = "{Total positivity, Grassmannians, and networks}",
      journal = {arXiv Mathematics e-prints},
         year = 2006,
        month = sep,
          eid = {math/0609764},
        pages = {math/0609764},
archivePrefix = {arXiv},
       eprint = {math/0609764},
 primaryClass = {math.CO},
       adsurl = {https://ui.adsabs.harvard.edu/abs/2006math......9764P}
}

@article{BernKLT,
    author = "Bern, Z. and Dixon, Lance J. and Perelstein, M. and Rozowsky, J.S.",
    title = "{Multileg one loop gravity amplitudes from gauge theory}",
    eprint = "hep-th/9811140",
    archivePrefix = "arXiv",
    reportNumber = "SLAC-PUB-7992, UCLA-98-TEP-28, UFIFT-HEP-98-33",
    doi = "10.1016/S0550-3213(99)00029-2",
    journal = "Nucl. Phys. B",
    volume = "546",
    pages = "423--479",
    year = "1999"
}

@article{KLT,
      author         = "Kawai, H. and Lewellen, D. C. and Tye, S. H. H.",
      title          = "{A Relation Between Tree Amplitudes of Closed and Open
                        Strings}",
      journal        = "Nucl. Phys.",
      volume         = "B269",
      year           = "1986",
      pages          = "1-23",
      doi            = "10.1016/0550-3213(86)90362-7",
      reportNumber   = "CLNS-85/667",
      SLACcitation   = "%%CITATION = NUPHA,B269,1;%%"
}

@article{Cachazo:2013gna,
      author         = "Cachazo, Freddy and He, Song and Yuan, Ellis Ye",
      title          = "{Scattering equations and Kawai-Lewellen-Tye
                        orthogonality}",
      journal        = "Phys. Rev.",
      volume         = "D90",
      year           = "2014",
      number         = "6",
      pages          = "065001",
      doi            = "10.1103/PhysRevD.90.065001",
      eprint         = "1306.6575",
      archivePrefix  = "arXiv",
      primaryClass   = "hep-th",
      SLACcitation   = "%%CITATION = ARXIV:1306.6575;%%"
}

@article{Cachazo:2013hca,
      author         = "Cachazo, Freddy and He, Song and Yuan, Ellis Ye",
      title          = "{Scattering of Massless Particles in Arbitrary
                        Dimensions}",
      journal        = "Phys. Rev. Lett.",
      volume         = "113",
      year           = "2014",
      number         = "17",
      pages          = "171601",
      doi            = "10.1103/PhysRevLett.113.171601",
      eprint         = "1307.2199",
      archivePrefix  = "arXiv",
      primaryClass   = "hep-th",
      SLACcitation   = "%%CITATION = ARXIV:1307.2199;%%"
}

@article{Cachazo:2013iea,
      author         = "Cachazo, Freddy and He, Song and Yuan, Ellis Ye",
      title          = "{Scattering of Massless Particles: Scalars, Gluons and
                        Gravitons}",
      journal        = "JHEP",
      volume         = "07",
      year           = "2014",
      pages          = "033",
      doi            = "10.1007/JHEP07(2014)033",
      eprint         = "1309.0885",
      archivePrefix  = "arXiv",
      primaryClass   = "hep-th",
      SLACcitation   = "%%CITATION = ARXIV:1309.0885;%%"
}

@phdthesis{Mizera:2019gea,
    author = "Mizera, Sebastian",
    title = "{Aspects of Scattering Amplitudes and Moduli Space Localization}",
    eprint = "1906.02099",
    archivePrefix = "arXiv",
    primaryClass = "hep-th",
    doi = "10.1007/978-3-030-53010-5",
    school = "Princeton, Inst. Advanced Study",
    year = "2020"
}

@article{ManganoParke,
    author = "Mangano, Michelangelo L. and Parke, Stephen J.",
    title = "{Multiparton Amplitudes in Gauge Theories}",
    eprint = "hep-th/0509223",
    archivePrefix = "arXiv",
    reportNumber = "FERMILAB-PUB-90-113-T",
    doi = "10.1016/0370-1573(91)90091-Y",
    journal = "Phys. Rept.",
    volume = "200",
    pages = "301--367",
    year = "1991"
}

\end{document}